%% file: main.tex
\documentclass[sigconf,authorversion,nonacm,table]{acmart}

\AtBeginDocument{%
  \providecommand\BibTeX{{%
    \normalfont B\kern-0.5em{\scshape i\kern-0.25em b}\kern-0.8em\TeX}}}

\usepackage{amsmath,amsfonts}
\usepackage{algorithm}
\usepackage{adjustbox}
\usepackage[noend]{algpseudocode}
\usepackage{textcomp}
\usepackage{soul}
\usepackage{booktabs}
\usepackage[utf8]{inputenc}
\usepackage[T1]{fontenc}
\usepackage{microtype}
\usepackage{rotating}
\usepackage{graphicx}
\usepackage{paralist}
\usepackage{tabularx}
\usepackage{multicol}
\usepackage{multirow}
\usepackage{pbox}
\usepackage{enumitem}	
\usepackage{colortbl}
\usepackage{pifont}
\usepackage{xspace}
\usepackage{url}
\usepackage{tikz}
\usepackage{tabularx}
\usepackage{fontawesome}
\usepackage{lscape}
\usepackage{listings}
\usepackage{color}
\usepackage{anyfontsize}
\usepackage{comment}
\usepackage{soul}
\usepackage{multibib}
\usepackage{multirow}
\usepackage{xspace}
\usepackage{caption}
\usepackage{subcaption}
\usepackage{footnote}
\usepackage{tcolorbox}
\usepackage{booktabs}
\usepackage{fixltx2e}
\usepackage{balance}
\usepackage[normalem]{ulem}

\definecolor{tabGreen}{rgb}{0,0,0}
\definecolor{tabRed}{rgb}{0,0,0}
\definecolor{tabBlue}{rgb}{0,0,0}

\newcommand{\cmark}{\text{\textcolor{tabGreen}{\ding{51}}}}
\newcommand{\xmark}{\text{\textcolor{tabRed}{\ding{55}}}}
\newcommand{\dmark}{\text{\textcolor{tabBlue}{\textemdash}}}

\newcommand\cval[1]{\text{\textcolor{black}{#1\%}}}

\newcommand*\circled[1]{\tikz[baseline=(char.base)]{
            \node[shape=circle,fill,inner sep=0.8pt] (char) {\textcolor{white}{#1}};}}

\definecolor{Gray}{gray}{0.9}
\definecolor{codegreen}{rgb}{0,0.6,0}
\definecolor{codegray}{rgb}{0.73,0.38,0.06}
\definecolor{codepurple}{rgb}{0.70,0.27,0}
\definecolor{codemagenta}{rgb}{0.74,0.09,0.42}
\definecolor{codeoutput}{rgb}{0.5,0,0}
\definecolor{backcolour}{rgb}{0.96,0.96,0.96}

\newboolean{showcomments}

\setboolean{showcomments}{true}

\ifthenelse{\boolean{showcomments}}
  {\newcommand{\nb}[2]{
    \fbox{\bfseries\sffamily\scriptsize#1}
    {\sf\small$\blacktriangleright$\textit{#2}$\blacktriangleleft$}
   }
   
  }
  {\newcommand{\nb}[2]{}
   
  }

\newcommand{\ie}{\emph{i.e.,}\xspace}
\newcommand{\eg}{\emph{e.g.,}\xspace}
\newcommand{\etc}{etc.\xspace}
\newcommand{\etal}{\emph{et~al.}\xspace}
\newcommand{\secref}[1]{Section~\ref{#1}\xspace}
\newcommand{\figref}[1]{Fig.~\ref{#1}\xspace}

\newcommand{\tabref}[1]{Table~\ref{#1}\xspace}

\newcommand{\tool}{{LANCE}\xspace}

\title{Using Deep Learning to Generate Complete Log Statements}



\begin{document}
	
\input{abstract}

\author{Antonio Mastropaolo}
\email{antonio.mastropaolo@usi.ch}
\affiliation{%
	\institution{SEART @ Software Institute Universit\`{a} della Svizzera italiana}
	\country{Switzerland}
}

\author{Luca Pascarella}
\email{luca.pascarella@usi.ch}
\affiliation{%
	\institution{SEART @ Software Institute Universit\`{a} della Svizzera italiana}
	\country{Switzerland}}

\author{Gabriele Bavota}
\email{gabriele.bavota@usi.ch}
\affiliation{%
	\institution{SEART @ Software Institute Universit\`{a} della Svizzera italiana}
	\country{Switzerland}}

\settopmatter{printfolios=true}
\maketitle

\input{introduction}

\input{approach}

\input{design}
\input{results}

\input{threats}
\input{related}

\input{conclusion}

\section*{Acknowledgment}

This project has received funding from the European Research Council (ERC) under the European Union's Horizon 2020 research and innovation programme (grant agreement No. 851720).

\bibliographystyle{ACM-Reference-Format}
\bibliography{main}

\end{document}

%% file: abstract.tex

\begin{abstract}

Logging is a practice widely adopted in several phases of the software lifecycle. For example, during software development log statements allow engineers to verify and debug the system by exposing fine-grained information of the running software. While the benefits of logging are undisputed, taking proper decisions about \emph{where} to inject log statements, \emph{what} information to log, and at which \emph{log level} (\eg error, warning) is crucial for the logging effectiveness. In this paper, we present \tool (\textbf{L}og st\textbf{A}teme\textbf{N}t re\textbf{C}omm\textbf{E}nder), the first approach supporting developers in all these decisions. \tool features a Text-To-Text-Transfer-Transformer (T5) model that has been trained on 6,894,456 Java methods. \tool takes as input a Java method and injects in it a full log statement, including a human-comprehensible logging message and properly choosing the needed log level and the statement location. Our results show that \tool is able to (i) properly identify the location in the code where to inject the statement in 65.9\% of Java methods requiring it; (ii) selecting the proper log level in 66.2\% of cases; and (iii) generate a completely correct log statement including a meaningful logging message in 15.2\% of cases.
\end{abstract}

%% file: introduction.tex

\section{Introduction} \label{sec:introduction}

Inspecting log messages is a popular practice that helps developers in several software maintenance activities such as testing~\cite{chen2018automated,chen2019experience}, debugging~\cite{satyanarayanan1992transparent}, diagnosis~\cite{zhou2019latent,yuan2012improving}, and monitoring~\cite{hasselbring2020kieker,harty2021logging}. Developers insert log statements to expose and register information about the internal behavior of a software artifact in a human-comprehensible fashion~\cite{oliner2012advances}. The data generated is used for runtime and post-mortem analyses. For example, when debugging log statements can support root cause analysis~\cite{lu2017log,gurumdimma2016crude}, while once the software is deployed logs can be used for performance monitoring~\cite{yao2018log4perf} or anomaly detection~\cite{meng2019loganomaly,zhang2019robust,du2017deeplog}.

Although technically possible, logging everything (\eg every exception) is inefficient and impracticable~\cite{zhu2015learning}. On the one hand, a coarse-grained logging risks hiding runtime failures, missing log messages useful for diagnoses~\cite{yuan2010sherlog}. 

\eject

On the other hand, a fine-grained logging risks increasing the overhead of log management and analysis~\cite{li2020qualitative}. To optimize the quantity and quality of data generated, developers insert log statements in strategic positions, specify appropriate log levels (\eg error, debug, info), and define compact but comprehensible text messages. Nonetheless, it remains a non-trivial task for developers to decide where, what, and at which level to log \cite{li2020qualitative,yuan2012characterizing}.

For these reasons, researchers have proposed techniques to support developers in deciding what parts of the system to log~\cite{yuan2010sherlog}, the log level for logging statements~\cite{li2020qualitative,yuan2012characterizing,oliner2012advances,li2017log}, and the structure of log messages~\cite{li2020towards}. For example, Jia \etal~\cite{jia2018smartlog} proposed an approach based on association rules to place error logs after code branches such as \emph{if} statements. Li \etal~\cite{li2018studying} studied the use of topic modeling for log placement at method-level. 
Also, two recent works tackled challenges related to log statements writing by adopting deep learning (DL) models. 

Li \etal~\cite{li2021deeplv}, with \emph{DeepLV}, pushed the boundaries of log recommendation by suggesting and fixing log levels of already typed log statements. \emph{DeepLV} relies on an ad-hoc DL network that combines syntactic (\ie AST) and contextual (\ie log message) information extracted from code to suggest an alternative logging level when needed. Li \etal~\cite{li2020shall} also proposed a second DL-based approach to provide fine-grained (\ie at the code block level) suggestions about where to add logging statements. The model captures both syntactic and semantic information of the source code and returns a binary value indicating whether to add or not a log statement in a given block. While achieving great performance, these techniques only partially support developers in logging practices. Indeed, none of them can generate complete log statements providing to the developer (i) the location where to inject it, (ii) the correct log level to use, and (iii) the actual log statement also featuring the needed natural language log message.

In this paper, we present \tool (\textbf{L}og st\textbf{A}teme\textbf{N}t re\textbf{C}omm\textbf{E}nder), an approach aimed at exploiting the recently proposed Text-To-Text-Transfer-Transformer (T5) model \cite{raffel2019exploring} to automatically generate and inject complete logging statements in Java code. We started by pre-training our model on a set of 6,832,859 Java methods. 
The pre-training has been performed through two pre-training objectives. The first is a classic ``masked token'' objective, in which we randomly mask 15\% of the code tokens in the Java methods asking the model to guess them. This provides the model with general knowledge about the Java language, including logging statements. 

\eject

The second pre-training objective provides as input to the model a Java method from which log statements originally present in it have been removed, with the T5 in charge of guessing where a log statement is needed by adding a special \texttt{$<$LOG\_STMT$>$} token. This provides the model with additional knowledge about \emph{where} logging is needed. Once pre-trained, the model is fine-tuned to generate complete log statements. In particular, given a Java method as input to \tool, we ask it to inject a complete log statement where needed. 
This means that \tool must generate a complete log statement and inject it in the proper location. 

In our evaluation, we asked \tool to automatically generate 12,020 log statements and compared them to the ones manually written by developers. We found that \tool is able to (i) correctly predict the appropriate location of a log statement in 65.9\% of cases; (ii) select a proper log level for the statement in 66.2\% of cases; and (iii) generate a completely correct logging statement, including a meaningful natural language message in 15.2\% of cases. Besides such a quantitative analysis we report and discuss qualitative examples of correct and wrong predictions generated by the model.

\textbf{Significance of research contribution.} Our work represents a step ahead in the automated support provided to developers for logging activities. Indeed, \tool is the first technique able to generate complete logging statements and to inject them in the right code location. \tool is complementary to techniques suggesting which parts of the system to log~\cite{yuan2010sherlog}, since it assumes that the method provided as input always needs a logging statement. In other words, we do not tackle the problem of deciding \emph{whether} a code component (in our case, a Java method) needs a logging statement, but we assume that such a decision has already been taken by another approach or by the developer itself. \tool can then take care of injecting the needed logging statements. 

We publicly release the code implementing our model and all data and additional scripts used in our study in a comprehensive replication package \cite{replication}.


%% file: approach.tex

\newcommand{\fd}{\emph{fine-tuning dataset}\xspace}
\newcommand{\pd}{\emph{pre-training dataset}\xspace}
\section{\tool: Log stAtemeNt reCommEnder} \label{sec:approach}
We start by overviewing the T5 model that is at the core of \tool and by explaining how we exploited it for the automation of logging activities (\secref{sub:t5}). Then, we describe the process used to build the datasets needed for its training, hyperparameter tuning and evaluation (\secref{sub:datasets}). \secref{sub:training} details the training of the model and the hyperparameter tuning we performed to identify the best configuration to use in our experiments. Finally, \secref{sub:predictions} explains how predictions are generated once the model has been trained. 

\subsection{Text-To-Text-Transfer-Transformer (T5)} \label{sub:t5}
Raffel \etal presented T5 \cite{raffel2019exploring} as a model that can be used to tackle any Natural Language Processing (NLP) task that can be expressed in a text-to-text format. This basically means that both the input and the output of the model are text strings. A single T5 model can be trained to support multiple tasks, such as machine translation (\eg from English to Franch) and question answering. The T5 demonstrated state-of-the-art performance on several NLP benchmarks~\cite{raffel2019exploring}. Also, it has been successfully used to automate code-related tasks \cite{mastropaolo2021studying}.

We do not discuss all the architectural details of T5, that are documented in \cite{raffel2019exploring}. However, it is worth mentioning that, as the name suggest, the T5 is a Transformer \cite{attention} model exploiting attention layers to weight the significance of the different parts composing the input strings. This is particularly useful when dealing with code-related tasks, since T5 can detect hidden and long-ranged dependencies among tokens, without assuming that nearest tokens are more related than distant ones. 

For example, the tokens representing the declaration of a local variable in the first statement of a method are related to the tokens implementing a return statement in which the value of such a variable is returned (despite the fact that the two statements could be far apart).

In our work, we exploit the specific architecture referred by Raffel \etal \cite{raffel2019exploring} as T5$_{small}$. Indeed, the authors present different versions of the T5 (small, base, large, 3 Billion, and 11 Billion) differing in complexity, size, and, consequently, training time. The T5$_{small}$ we adopted features a total of 60M parameters allowing reasonable training times with the hardware resources at our disposal. The code implementing the T5 model is available in our replication package \cite{replication}.

\subsubsection{Instantiating T5 to Automate Logging Activities}
\begin{figure}[h]
	\includegraphics[width=\linewidth]{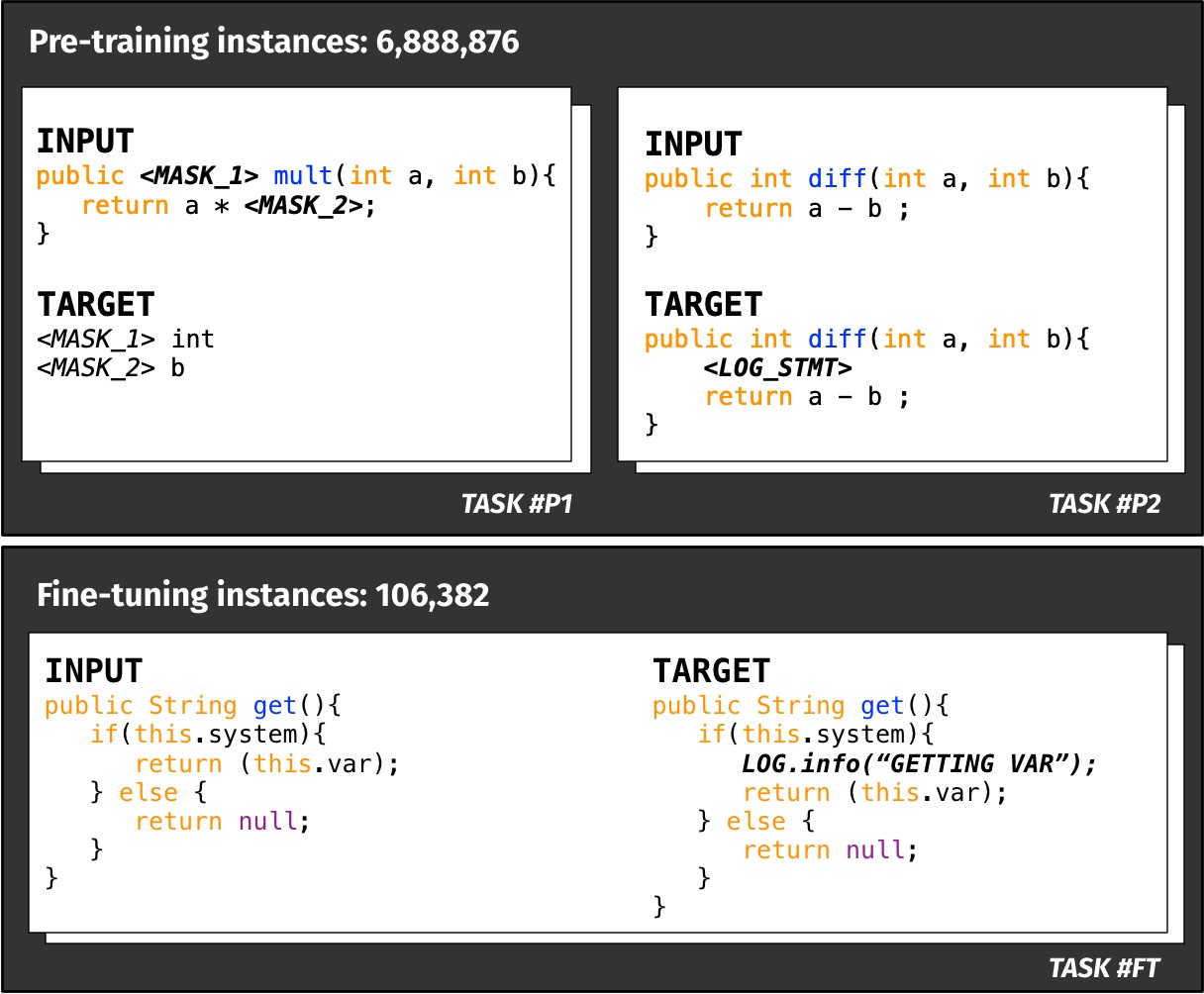}
	\caption{Pre-training and fine-tuning tasks}
	\label{fig:data}
\end{figure}

The T5 model is trained in two phases (i) pre-training, which allows defining a shared knowledge-base useful for a large class of text-to-text tasks (\eg guessing masked words in English sentences to learn about the language); and (ii) fine-tuning, which specializes the model on specific downstream tasks (\eg machine translation task). Both pre-training and fine-tuning can be performed in a multi-task setting (\ie a single model is trained on several tasks). \figref{fig:data} depicts the tasks we adopt for the T5 pre-training and fine-tuning, while the building of the needed datasets is detailed in \secref{sub:datasets}.

\eject

\textbf{Pre-training.} We targeted a two-fold goal for the T5 pre-training: (i) provide the model with knowledge about the underlying patterns of the Java language; and (ii) allow the model to learn where a log statement is needed in a given Java method, without focusing at this stage on the log level and the message to print. Concerning the first point, our first pre-training task (Task \#P1 in \figref{fig:data}) is a classic denoising task \cite{raffel2019exploring} in which we randomly mask $15\%$ of code tokens in each instance (\ie a Java method) asking the model to guess the masked tokens.

The second task (Task \#P2 in \figref{fig:data}), instead, asks the model to predict the correct position of a log statement within the input method. 
Basically, we provide as input to the model a Java method that originally had $n$ log statements ($n \geq 1$), with $n-1$ log statements (\ie we completely remove one log statement). Then, we ask the model to predict where the removed log statement was and to inject in that position a special $<$\texttt{LOG\_STMT}$>$ tag. This means that a method having $n$ log statements will appear $n$ times in the pre-training dataset, each time with a different log statement removed.

\textbf{Fine-tuning.} Once the model is pre-trained, the fine-tuning task (Task \#FT in \figref{fig:data}) specializes the model to the specific problem we target, namely the injection of complete log statements. Also in this case, the input is represented by a Java method that originally had $n$ log statements ($n \geq 1$) from which we completely remove one log statement (\ie same input as Task \#P2). However, as output, we expect the model to inject the actual log statement in the right position, choosing the right log level and a meaningful log message.


\subsection{Building the Training Datasets} \label{sub:datasets}
We detail the process used to build both the pre-training and the fine-tuning dataset. We mined Java projects on GitHub~\cite{github} by leveraging the search tool by Dabic \etal~\cite{dabic2021sampling}. The querying user interface~\cite{ghs} allows to identify GitHub projects that meet specific selection criteria. We selected all Java projects having at least 500 commits, 10 contributors, 10 stars, and not being forks (to reduce the chance of mining duplicated code). The commits/contributors/stars filters aim at discarding personal/toy projects. Instead, the decision of only focusing on Java projects simplifies the toolchain needed for our study and allows to train the model on a coherent code corpus.

This process resulted in 5,473 candidate projects. We successfully cloned the latest snapshot of 5,459 of them (some projects cannot be cloned since they were deleted or made private). Then, to foster even more the cohesiveness of our dataset, we decided to select only projects declaring a dependency towards Apache Log4j \cite{log4j}, a well-known Java logging library. To identify these projects, we firstly checked whether a \texttt{POM} (Project Object Model) file\footnote{A POM file is used in Maven to declare dependencies towards Maven libraries.} was present in the project's directory. If this was the case, we parsed it to check whether it featured a Log4j dependency. If no POM file or no Log4j dependency was found, the project was discarded. We found 1,465 Java projects having an explicit dependency towards Log4j.

We then used \emph{srcML}~\cite{SrcML} to extract from these projects all their Java methods ($\sim$10M).  Of these methods, $\sim$96\% do not have log statements. This still provides us with $\sim$320k methods featuring at least one log statement. Then, we filtered out all methods having  \textit{\#tokens $\geq$ 512} or \textit{\#tokens $<$ 10}, where \textit{\#tokens} represents the number of tokens composing a method (excluding comments). 

The filter on the maximum number of tokens is needed to limit the computational expense of training DL-based models (similar choices have been made in previous works \cite{Tufano:tosem2019,haque:2020,tufano2021automating}). Finally, we remove duplicate methods from the dataset to avoid overlap between training and test sets we built from them. This left us with 6,909,280 methods, that we used to create the pre-training and the fine-tuning datasets summarized in \tabref{tab:datasets}.

\begin{table}[h!]
	\footnotesize
		\caption{Num. of methods in the datasets used in our study}
		\begin{tabular}{lrrcrcr}
				\toprule
				\multirow{2}{*}{\textbf{Dataset}} & \multicolumn{2}{c}{\textbf{train}} && \textbf{eval}     && \textbf{test} \\ \cline{2-3} \cline{5-5} \cline{7-7}
				& \textbf{w/ log} & \textbf{w/a log} && \textbf{w/ log} && \textbf{w/ log} \\\midrule
				\pd &  & & & & &\\
				\hspace{6mm}\emph{$Task_{\#P1}$} & -  & 6,755,884 &  &- &  &- \\
		     	\hspace{6mm}\emph{$Task_{\#P2}$}& 76,975  & -& & -&  &- \\
				\fd                           & 61,597            & - & &7,699  & & 7,125           \\
				\bottomrule
		\end{tabular}
	\label{tab:datasets}
\end{table}

We used all the methods not having log statement (w/a log in \tabref{tab:datasets}) to build the dataset needed for the pre-training Task \#P1 (\ie the denoising task in which we randomly mask 15\% of tokens). For the pre-training Task \#P2 (\ie guessing the correct position of a log statement), we used 50\% of methods with log statements: A method featuring $n$ log statements is present $n$ times in the dataset, each time with a different log statement removed.

The remaining 50\% of methods featuring a log statement have been used instead for building the fine-tuning dataset. The latter has been split into 80\% training, 10\% evaluation, and 10\% test. The evaluation has been used to perform the hyperparameter tuning of the model (\secref{sub:training}), while the test set represents the instances on which the performance of \tool have been assessed (\ie its ability to generate correct log statements in the right location).

\subsection{Model Training and Hyperparameter Tuning} \label{sub:training}
The pre-training has been performed for 300k steps. We used a 2x2 TPU topology (8 cores) from Google Colab to pre-train the model with a batch size of 128. As a learning rate, we use the \emph{Inverse Square Root} with the canonical configuration \cite{raffel2019exploring}. We also used the pre-training dataset and 3,244,116 English sentences coming from the \emph{C4} dataset \cite{raffel2019exploring} to train a \emph{SentencePiece} model (\ie a tokenizer for neural text processing). We decided to train the tokenizer on both Java code and English natural language to make sure it can deal with complex log messages. We set its size to 32k word pieces.

Once pre-trained, the model can be fine-tuned. However, before that, we performed the same hyperparameters tuning used by Mastropaolo \etal \cite{mastropaolo2021studying} when employing the T5 for code-related tasks: We do not tune the hyperparameters of the T5 model for the pre-training (using the default ones), but we experiment with four different learning rates, namely constant learning rate (C-LR), slanted triangular learning rate (ST-LR), inverse square learning rate (ISQ-LR), and polynomial learning rate (PD-LR). Our replication package \cite{replication} reports the exact setting used for each of the experimented learning rates (\eg the constant learning rate was set to 0.001, \etc).

Before detailing the hyperparameter tuning, we must anticipate that in our study (\secref{sec:design}) we assess the performance of \tool in four different pre-training scenarios. 

First, to verify the impact on performance of the pre-training phase, we perform an ablation study in which the model is not pre-trained, but directly fine-tuned (\emph{No pre-trained} scenario). Second, we pre-train a model by employing a multi-task pre-training setting in which both our pre-training tasks (\ie Task \#P1 and \#P2 in \figref{fig:data}) are used (\emph{Multi-task}). Finally, we assess the performance of T5 when pre-trained by only using Task \#P1 (\emph{Denoising-task}) or Task \#P2 (\emph{LogStmt-task}). 

Once pre-trained, all models are fine-tuned and compared. This allows to assess the contribution to performance (if any) brought by the different pre-training strategies. 

Having four different training scenarios and four possible learning rates, the hyperparameter tuning required building 16 models. We fine-tuned each model (\ie each configuration) for 100k steps. Then, we compute the percentage of correct predictions (\ie cases in which the model can inject a correct log statement in the right position) achieved in the evaluation set. The achieved results are reported in \tabref{tab:hyperparameter_results}. The best configuration of each scenario (in boldface) is the one that has been used in our study to assess the \tool's performance after fine-tuning the models for 200k steps. 

\begin{table}[h!]
	\centering
	\caption{T5 hyperparameter tuning results}
		\begin{tabular}{lrrrr}
			\toprule
			\textbf{Experiment}                  										& \textbf{C-LR}              & \textbf{ST-LR}      & \textbf{ISQ-LR}        & \textbf{PD-LR} \\
			\midrule
	 	    \emph{Multi-Task}                         &   11.84\%                & 11.88\%    		           & \textbf{12.30}\%           &  12.18\%         \\
			\emph{LogStmt-Task}                    &  11.76\%                & 11.74\%      	   				&  \textbf{12.36\%}                  					&  11.70\%         \\	
			\emph{Denoising-Task}                    &   15.01\%                & 13.62\%      	             &  14.80\%                 	 &  \textbf{15.12\%}         \\		
			\emph{No Pre-training}                        	     	                              &   12.64\%                & \textbf{13.25\%}      	   &  13.12\%                  &  12.65\%         \\			
			\bottomrule
		\end{tabular}
	
	\label{tab:hyperparameter_results}
\end{table}

%

\subsection{Generating Predictions} \label{sub:predictions}
Once the T5 model is trained, it can generate predictions using different decoding strategies. For this first work on log statement generation, we decided to target the \emph{greedy decoding} strategy: Predictions are generated by selecting the token with the highest probability of appearing in a specific position at each time step. This means that for a given input, a single prediction is generated (\ie the one considered the most likely by the model).

%% file: design.tex

\section{Study Design} \label{sec:design}

The \textit{goal} of this study is to evaluate the performance of \tool in automatically generating and injecting complete logging statements in Java methods. The \textit{context} is represented by the datasets described in \secref{sec:approach}. 

We aim at answering the following research questions (RQs):

\begin{itemize}

\item[\textbf{RQ$_1$:}]\textit{To what extent is \tool able to correctly inject complete logging statements in Java methods?} With RQ$_1$ we aim at assessing the performance of the trained T5 models in generating and injecting in the correct position logging statements in unseen Java methods. Besides quantitatively answer this RQ by reporting the percentage of correct predictions generated by \tool, we manually inspected the generated log statements to discuss interesting cases of correct and wrong predictions.
\smallskip

\item[\textbf{RQ$_2$:}]\textit{How do different pre-training strategies impact the performance of \tool?} RQ$_2$ analyzes the impact of different pre-training strategies on \tool's performance. In particular, we experiment with the four T5 variants described in \secref{sub:training}: \emph{No Pre-training}, \emph{Multi-Task}, \emph{LogStmt-Task}, and \emph{Denoising-Task}.


\end{itemize}

\subsection{Data Collection and Analysis} \label{sub:analysis}

To answer RQ$_1$ and RQ$_2$ we run against the test set (\tabref{tab:datasets}) the best-performing configuration (\secref{sub:training}) of the four models output of the different pre-trainings. 

Then, we assess the accuracy of the predictions generated by each model. In this regard, we rely on the code (\ie log statements) manually written by developers as a ground truth. This is a common practice \cite{haque:2020, li2021deeplv, Tufano:tosem2019, tufano2021automating} concerning the definition of the oracle (\ie the output the model is expected to generate). Hence, first, we compute the percentage of correct predictions, namely cases in which \tool correctly synthesizes the log statement (\ie both the log level and the log message were correct) while injecting it in the correct position in the method (\ie the same position adopted by developers). Successively, we assessed the extent to which \tool generates ``partially correct'' predictions. In particular, there are three important ``components'' that the T5 model has to predict when it comes to the addition of a log statement: its level, message, and position (location in the method). We compute the percentage of cases in which \tool was able to correctly predict (i) at least one of these three ``components'' (\eg the log level is correct, but the message as well as the position are different from the reference); and (ii) at least two of the ``components'' (\eg the log level and message are correct, but the statement is injected in the wrong position). The third scenario (\ie all three ``components'' are correct) is represented by the previously discussed correct predictions. The number of instances in the test set is 12,020, where each instance represents one of the 7,125 Java methods in \tabref{tab:datasets} with a specific log statement removed (one method can have multiple log statements). \smallskip


\textbf{Manual analysis.} 
On top of this quantitative analysis, we also performed qualitative analyses aimed at better understanding the strengths and weaknesses of \tool. Besides reporting interesting cases of perfect predictions generated by our approach, we manually inspected a set of ``wrong predictions'' generated by the model. In particular, we focused on wrong predictions in which the log level and the location were correct. This means that the difference between the generated and the target log statement was the natural language message. Such a decision was driven by the goal of our manual analysis, aimed at understanding whether the generated log, while different, represented a good alternative to the reference one. This is unlikely to happen if the log level or location is different from the target or, at least, it is tough to judge for people not directly involved in the development of the code in which the log statement is injected (such as the authors who inspected these wrong predictions). For these reasons, we randomly selected 300 wrong predictions in which, however, the log level and the log location were correctly guessed by the model. Then, two authors manually inspected the original method with the log statement written by the developers and the same method with the statement generated by \tool to classify it in one of the following categories:

\begin{enumerate}
\item \emph{Same information}: The generated log statement is semantically equivalent to the target one. This could happen in the case in which the log messages of the two statements express the same information but with different wordings (\eg ``\emph{exception thrown when invoking method getPaper()}'' \emph{vs} ``\emph{getPaper() thrown an exception}''). It even happens, as we will show, that \tool proposes a more expressive log message as compared to the one used by developers.

\item \emph{Meaningful}: The generated log statement includes an articulated message that can be understood, but it is not equivalent to the one present in the target log statement.


\item \emph{Meaningless}: The generated log statement includes a message that is meaningless considering the context (\ie the method in which it has been injected) and/or the logging message cannot be comprehended.
\end{enumerate}

The two authors analyzed each of the 300 statements independently from each other, agreeing on the classification of 189 (63\%) log statements. The remaining 111 ($\sim$37\%), representing conflicts, have been solved by a third author not involved in the first stage. 

The 300 instances to manually validate have been selected from the predictions generated by the best-performing model. Such a qualitative analysis is fundamental in our study since the quantitative metrics we adopt consider as wrong predictions also cases in which the predicted and the reference log statements are very similar or equivalent (\eg they differ for one word in the log message).

%% file: results.tex

\section{Results Discussion} \label{sec:results}
To simplify the discussion of the results, we answer our research questions together through a quantitative and qualitative analysis.

\input{tables/results-predictions}

\subsection{Quantitative Analysis}

\tabref{tab:predictions} reports the results achieved by the four experimented models output of different pre-training strategies (\ie \textit{Multi-Task}, \textit{LogStmt-Task}, \textit{Denoising-Task}, and \textit{No pre-training}) in terms of correct predictions. \tabref{tab:predictions} shows the correct predictions for all combinations of the three ``components'' to predict (\ie log level, position where to inject it, and log message). In other words, we analyze cases in which (i) at least one of the three components to predict was correct (\eg at least the level), (ii) at least two were correct (\eg level and location), and (iii) the entire log statement was correctly synthesized (level, position, and message). 

\tabref{tab:predictions} can be read as follows. Each row includes three symbols below the three components to predict. The check mark (\cmark) below a component $c$ indicates that the results reported in that row refer to log statements in which $c$ was correctly predicted. The dash mark (\dmark), instead, indicates that $c$ can be either correct or wrong for the predictions in that row. Finally, the cross mark (\xmark) indicates that the component was wrongly predicted for the corresponding log statements. For example, the very first line:

\begin{quote}
\centering
	$Level=\cmark ~\land~ Position=\dmark ~\land~ Message=\dmark$
\end{quote}

\noindent indicates that the row reports the percentage of predictions in which the log level was correctly predicted, independently from the prediction of the position and message, which could be correct or wrong. Instead, the third row:

\begin{quote}
	$Level=\cmark ~\land~ Position=\xmark ~\land~ Message=\xmark$
\end{quote}

\noindent reports cases in which the level was correctly predicted, but the guessed position and the generated message were wrong. The row labeled with ``\emph{All}'' represents the most challenging scenario since it implies that the generated log statement was identical to the reference one in all its parts. Finally, the last black row shows the percentage and the absolute value of the instances in the test set that cannot be parsed with a \emph{JavaParser}~\cite{javaparser} due to syntax errors.

Concerning RQ$_2$, the best-performing model is the one pre-trained by using only the denoising task (\ie 15\% of masked tokens). 

Indeed, its performance is substantially superior to T5 pre-trained in a multi-task setting. For example, in terms of completely correct predictions, this model achieves a 15.20\% of ``perfect'' log statements versus the 12.32\% of the multi-task model. Also, the \emph{Denoising-task} model is the one generating the lowest number of syntactically wrong log statements (2.33\%). While such a result might seem surprising initially, a possible explanation could be the specific pre-training and fine-tuning we performed: Our fine-tuning task, requiring the model to generate and inject a complete log statement in a Java method, is quite similar to the \emph{LogStmt-Task} pre-training task. Indeed, the latter asks the model to inject a \texttt{<LOG>} placeholder in the position in which a log statement has been removed in a Java method. Thus, it is possible that such an additional pre-training did not benefit the model's performance. 

More in general, concerning the role played by the pre-training, we observed a substantial boost of performance only in the case of the \emph{Denoising-task} (+2.51\% of perfectly predicted log statements). The other two pre-trainings only marginally improved the performance of the base model (see \tabref{tab:predictions}). In the following, we focus on the best-performing model. 

\tool correctly predicts the log level in 66.24\% of cases, while the position in which a log statement should be injected is correctly guessed in 65.40\% of cases. This two-fold achievement (\ie log level and position) suggests that \tool can effectively support developers with logging activities. Looking at the third row in \tabref{tab:predictions}, it is instead clear that \tool struggles to generate logging messages that are identical to the ones manually written by developers (success in 16.90\% of cases). This difference in performance among the three log statement ``components'' to predict is kind of expected. Indeed, log level and the position have a quite small search space: the log level can assume one out of six possible values in Log4j (\emph{Trace}, \emph{Debug}, \emph{Info}, \emph{Warn}, \emph{Error}, and \emph{Fatal}), while the position in which a log statement can be injected is bounded to the number of tokens composing the methods, being at most 512. Instead, the natural language log message can be written in many different forms, reducing the chances of obtaining two identical sentences. 

In 35\% of cases, both the level and the position are correctly guessed with, however, a wrong logging message, while in an additional 15.20\% all three ``components'' are correctly synthesized. The generation of the logging message basically acts as a sort of upper bound for the performance of \tool, with 16.90\% of generated logs having a correct message and an overall 15.20\% of completely correct log statements (\ie level, position, and message are correct).

\begin{figure}[h]
	\includegraphics[width=\linewidth]{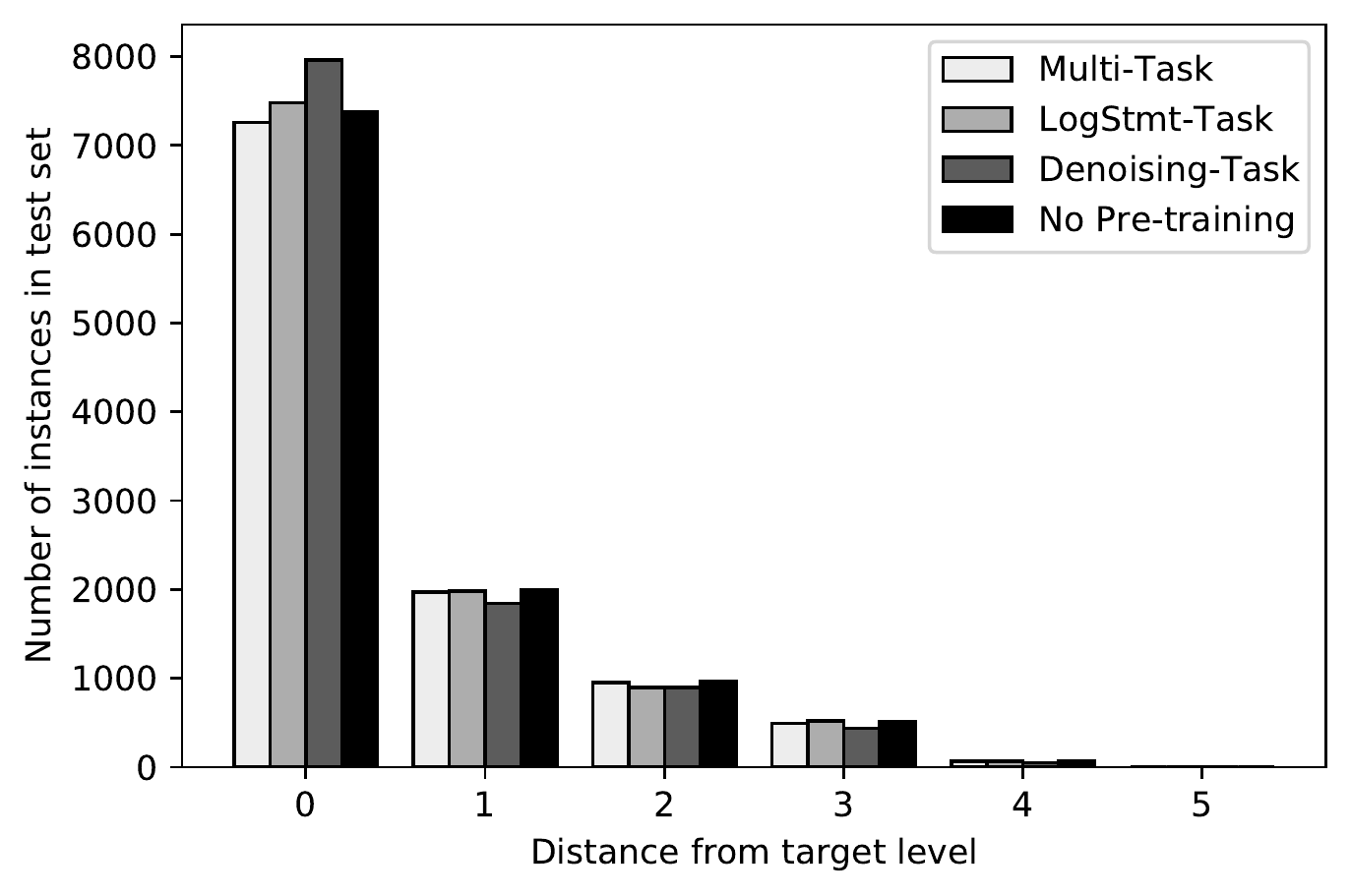}
	\caption{Log level distance in the predictions generated by \tool. Zero indicates predictions having a correct level.}
	\label{fig:histogram_level}
\end{figure}

\begin{figure}[h]
	\includegraphics[width=\linewidth]{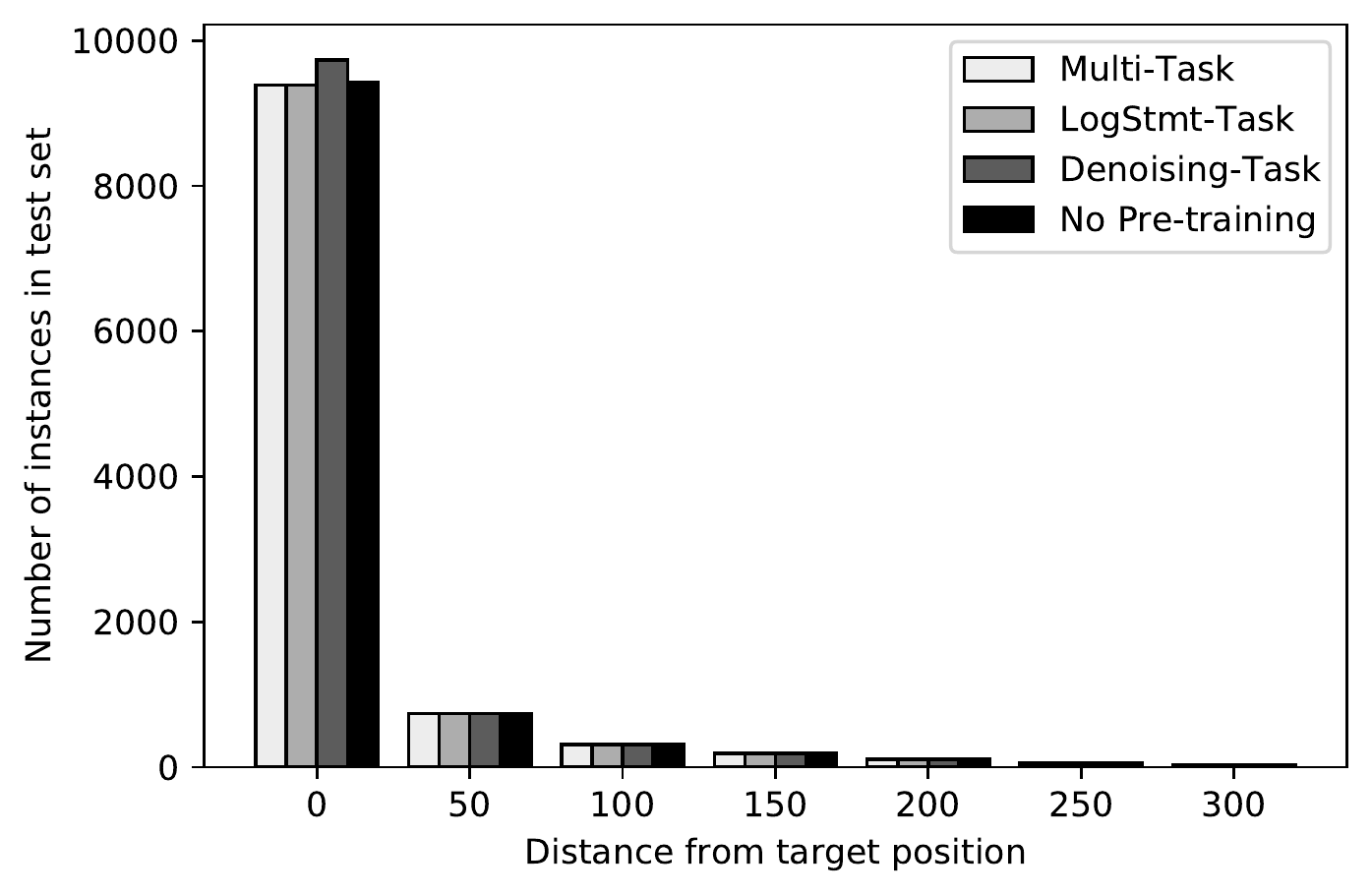}
	\caption{Distance in terms of \#Tokens between the position of the log statement in the prediction and in the target. Zero indicates predictions injected in the right position.}
	\label{fig:histogram_token}
\end{figure}

While \tool performs quite well in predicting the log level and position, there is still a large percentage of log statements for which their prediction fails. However, the ``magnitude'' of the error made in the prediction can substantially vary for both these tasks. 

Concerning the log level, the six levels defined in Log4j can be sorted based on their priority: (1) \emph{Trace}, (2) \emph{Debug}, (3) \emph{Info}, (4) \emph{Warn}, (5) \emph{Error}, and (6) \emph{Fatal}. For example, the \emph{Info} level is used for logging informational messages and tracking the behavior of the running software (\eg method $m$ starts the execution). In contrast, the last level, \emph{Fatal}, is designated for logging severe errors, or in other words, for logging the behavior likely to compromise the execution of the software (\eg by causing a crash). 

A wrong level prediction made by \tool could recommend the usage of \emph{Info} instead of \emph{Fatal} as well as the usage of \emph{Error} instead of \emph{Fatal}. However, these two errors have a different magnitude, with the first completely misleading the developer while the latter resulting in a sub-optimal (but still acceptable) log level decision. \figref{fig:histogram_level} depicts a histogram showing the number of instances in our test set for which the log level prediction had a distance from the target level going from 0 (\ie the level was correctly predicted) to 5 (\ie the worst-case scenario indicating a \emph{Trace} recommended instead of \emph{Fatal} or \emph{vice versa}). While we report the results achieved by all models, also in this case, we focus our discussion on the best-performing one (\emph{Denoising-task}). Note that not all 12,020 instances from our test set are depicted in \figref{fig:histogram_level}. Indeed, besides the ones containing syntax errors (281 for the \emph{Denoising-task}), we also had to exclude 557 instances for which the model did not recommend a valid log level, making impossible the computation of the distance.


As it can be seen, besides the instances with a correctly predicted level (7,962), most of the errors are just one level far from the target (1,842--16.5\%--instances), while very rarely the difference is higher than two (483--4.3\%--instances).

\figref{fig:histogram_token} depicts the same analysis performed, however, for the prediction of the location where to inject the log statement. In this case, the $x$-axis reports the distance (in code tokens) of the predicted location from the target location. 

The reported numbers must be interpreted as ``up to $n$ tokens far'' (\eg up to 50, up to 100, \etc). As it can be seen, the wrong locations are mostly in areas close to the target location: Besides the 7,890 correctly predicted, an additional 1,864 ($\sim$17\%) fall within 50 code tokens, while only 723 ($\sim$6\%) are more than 100 tokens far.

\input{tables/results-performance}

Finally, we investigated whether the performance of \tool is affected by the level of the log statement. Indeed, it is possible that specific types of log statements are easier to predict than others. 

\tabref{tab:logLevel} reports for each log level and for each of the experimented model variants: (i) the percentage of predictions in which the log level was correctly predicted independently from the prediction of the location and message (column ``All''); and (ii) the percentage of completely correct predictions (column ``Corr. Pred.''). 

A first observation that can be made from \tabref{tab:logLevel} is that \tool provides good and similar performance across all log levels. The highest percentage of correct predictions is for the \emph{Error} level, while the lowest is for the \emph{Warn} level. Indeed, focusing on the best-performing model, for 76.10\% of \emph{Error} instances in the test set the log level is correctly inferred, against the 60.20\% of the \emph{Warn} instances. Similarly, the percentage of completely correct predictions, moves from 23.40\% (\emph{Error}) to 9.59\% (\emph{Warn}). This finding is consistent across all models and we believe it is due to the simpler messages usually adopted in statements logging errors. To verify such a conjecture, we computed the number of characters composing log statements having different levels. We found that, on average, \emph{Error} instances are composed by $\sim$70 characters, against the $\sim$215 of \emph{Warn} instances. However, these numbers only tell part of the story. Indeed, we found that the \emph{Info} level, which is the second worst in terms of performance, features statements composed, on average, by $\sim$79 characters. Thus, despite being similar in size compared to the \emph{Error} instances, we still observed a drop of performance. We believe that this result is due to the fact that messages in \emph{Info} log statements are much more variegated as compared to error messages. Finally, it is worth highlighting the good performance achieved at the \emph{Debug} level, which points to the possible usage of techniques such as \tool in supporting bug localization activities by recommending the injection of proper debug statements. Clearly, additional performance improvements are needed before considering \tool ready to be adopted by developers.

The analysis of the ``wrong'' predictions is difficult to perform quantitatively for log messages as done for the level and the position. One possibility is to compute the BLEU (Bilingual Evaluation Understudy) score \cite{Papineni:2002} between the generated and the reference messages. BLEU is used to assess the quality of an automatically generated text. Such a score ranges between 0.0 and 1.0, with 1.0 indicating that the generated and the reference message are identical. We adopt the BLEU-4 variant, which computes the overlap in terms of 4-grams between the generated and the reference messages. 

Concerning the best-performing model (similar findings hold for the other models), we obtained an average BLEU-4 of 0.15. 

However, such a number is difficult to interpret since there is no accepted threshold above which an automatically generated text is considered of good quality. For this reason, we rely on the qualitative analysis introduced in \secref{sub:analysis} and discussed in the following.

\subsection{Qualitative Analysis}

Among the 300 ``wrong'' predictions analyzed, we found 85 of them (28.33\%) to report the \emph{same information} of the target predictions (\ie the log message was different but semantically equivalent); 209 (69.66\%) to include a \emph{meaningful} but not equivalent message; and 6 (2\%) to include \emph{meaningless} messages. Thus, in the set of predictions we considered ``wrong'' in our quantitative analysis due to the different log message generated as compared to the reference one, we can estimate a $\sim$28\% of predictions that are still valuable.

\begin{figure*}
	\includegraphics[width=0.8\linewidth]{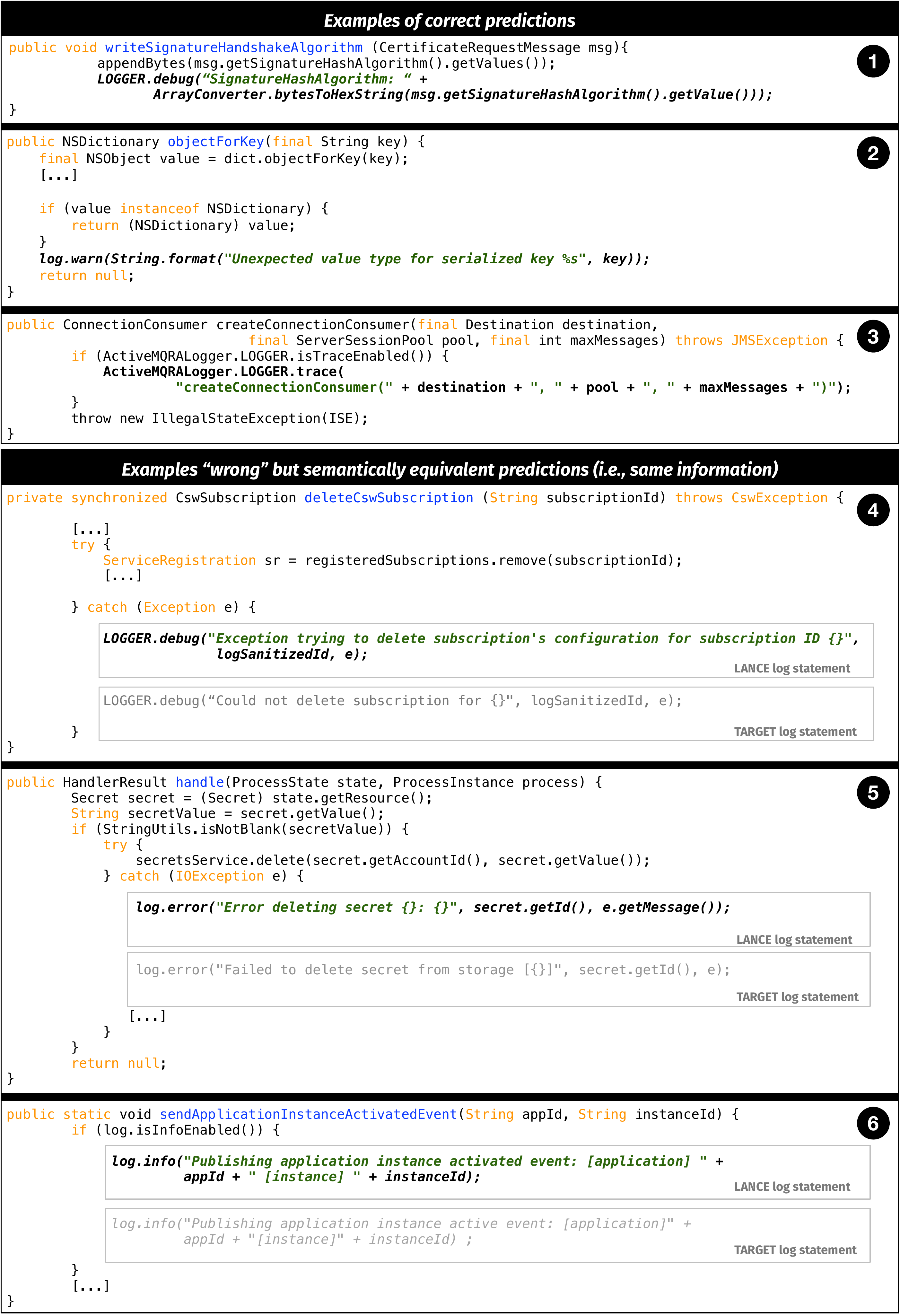}
			\caption{Examples of log statements generated by \tool}
			\label{fig:qualitative}
\end{figure*}

To better understand the capability of \tool in generating log statements, we report in \figref{fig:qualitative} three examples of (i) correct predictions (top part of the figure), in which the generated log statement is equivalent to the one written by developers in all its parts; and (ii) ``wrong predictions'' classified in our manual analysis as reporting the same information of the target prediction. We summarized the methods in order to only show statements that are relevant to understand the injected log statement (irrelevant statements are replaced by $[\dots]$).

Concerning the correct predictions, we just show the method with the generated log statement that, as said, is identical to the one written by the developers. In the first example \circled{1}, \tool injected a statement to log the state of the \texttt{msg} object. What it is interesting about this instance is that the model understood the need for invoking the method \texttt{Array\-Converter.\-bytes\-ToHex\-String} in order to obtain a human comprehensible representation of the logged state. 

In the second instance \circled{2}, \tool inferred that if the \texttt{if} condition is not satisfied (\ie \texttt{value} \texttt{instanceof} \texttt{NSDictionary}), this indicates an unexpected value type for the passed parameter (\texttt{key}). In other words, the model mapped the \texttt{instanceof} operator to possible issues related to object types.

The last correct prediction in \figref{fig:qualitative} \circled{3} shows instead the ability of \tool to compose log messages by using the appropriate syntax needed to concatenate several parameters to string elements. In this case, the log statement is just aimed at reporting when the execution of a specific method starts.

Moving to the ``wrong'' but semantically equivalent predictions (bottom part of \figref{fig:qualitative}), instance \circled{4} shows an interesting case in which the log message synthesized by \tool (\ie ``\emph{Exception trying to delete subscription's configuration for subscription ID \{\}}'') is even more detailed than the one written by developers (\ie ``\emph{Could not delete subscription for \{\}}). In \circled{5}, instead, the opposite occurs (\ie the manually written message is more detailed) with the two messages, however, communicating similar information. 

Finally, the last example \circled{6} shows two messages only differing for one word (\emph{activated} \emph{vs} \emph{active}). This example is representative of many instances we found in which differences were even smaller. For example, we observed cases in which the only difference was the usage of letter case. Indeed, in our quantitative analysis we decided to be conservative, considering a prediction as correct only if it matched the reference one even in terms of letter case. This was done to avoid considering as correct predictions making a wrong usage of camelCase.

%% file: tables/results-predictions.tex

\begin{table*}[ht]
	\caption{Correct predictions considering the three-dimensional challenges of injecting log statements.}
	\label{tab:predictions}
				\begin{tabular}{cccccrrrr}
					\toprule
											       				& \multicolumn{3}{c}{\textbf{Log}}                                                      			&& \multicolumn{4}{c}{\textbf{Predictions}}                                                                   						       																\\
											       				\cline{2-4} \cline{6-9} 
			                  									& \multirow{2}{*}{\textbf{Level}}	& \multirow{2}{*}{\textbf{Position}}& \multirow{2}{*}{\textbf{Message}} && \multicolumn{3}{c}{\textbf{Pre-training}}                              		        			& \multicolumn{1}{c}{\multirow{2}{*}{\textbf{No Pre-training}}} \\
			                									&                        			&                           &                          			&& \multicolumn{1}{c|}{\textit{Multi-Task}}	& \multicolumn{1}{c|}{\textit{LogStmt-Task}}	& \multicolumn{1}{c}{\textit{Denoising-Task}}		&																\\

                    \midrule \multirow{6}{*}{\rotatebox[origin=c]{90}{\textit{1 out of 3}}}	   
                    											& \cmark                   			& \dmark                    & \dmark                   			&& \cval{60.37}   							& \cval{62.22}    							   	& \cval{66.24}    									& \cval{61.39}						\\
                           	 									& \dmark                			& \cmark                    & \dmark                   			&& \cval{60.20}   							& \cval{60.70}    								& \cval{65.40}   									& \cval{60.36}						\\
                            									& \dmark                			& \dmark                    & \cmark                   			&& \cval{13.74}   							& \cval{13.85}    								& \cval{16.90}   									& \cval{14.37}						\\
																& \cmark                			& \xmark                    & \xmark                   			&& \cval{15.50}   							& \cval{16.27}    								& \cval{15.14}   									& \cval{16.09}						\\
                           	 									& \xmark                			& \cmark                    & \xmark                   			&& \cval{15.06}  							& \cval{14.45}    								& \cval{13.94}   									& \cval{14.71}						\\
                            									& \xmark                			& \xmark                    & \cmark                   			&& \cval{0.06}   							& \cval{0.10}    								& \cval{0.07}   									& \cval{0.05}						\\
                    \midrule
					\multirow{3}{*}{\rotatebox[origin=c]{90}{\textit{2 out of 3}}}	
					            								& \cmark                			& \cmark                    & \xmark                   			&& \cval{31.74}  							& \cval{32.79}    								& \cval{35.00}   									& \cval{31.73}						\\
                            									& \cmark                			& \xmark                    & \cmark                   			&& \cval{0.29}   							& \cval{0.29}    								& \cval{0.35}   									& \cval{0.39}						\\
                            									& \xmark                			& \cmark                    & \cmark                   			&& \cval{0.56}   							& \cval{0.60}    								& \cval{0.71}   									& \cval{0.75}						\\
                    \midrule
\multirow{2}{*}{\rotatebox[origin=c]{90}{\textit{All}}}			& \multirow{2}{*}{\cmark}			& \multirow{2}{*}{\cmark}	& \multirow{2}{*}{\cmark}			&& \multirow{2}{*}{\cval{12.32}}				& \multirow{2}{*}{\cval{12.33}}					& \multirow{2}{*}{\cval{15.20}}						& \multirow{2}{*}{\cval{12.69}}		\\
																&									&							&									&&											&												&													& 									\\
\rowcolor[HTML]{000000}											&									&							&									&& \color[HTML]{FFFFFF}{5.77\%}				& \color[HTML]{FFFFFF}{4.15\%}					& \color[HTML]{FFFFFF}{2.33\% }						& \color[HTML]{FFFFFF}{3.35\%}		\\
\rowcolor[HTML]{000000} \multirow{-2}{*}{\color[HTML]{FFFFFF}{Wrong Syntax}}&								&							&									&& \color[HTML]{FFFFFF}{(694/12,020)}			& \color[HTML]{FFFFFF}{(499/12,020)}				& \color[HTML]{FFFFFF}{(281/12,020)}					& \color[HTML]{FFFFFF}{(403/12,020)}	\\
					\bottomrule
				\end{tabular}
\end{table*}

%% file: tables/results-performance.tex

\begin{table*}[]
	\centering
	\caption{Percentage of correct predictions by log statement level. Column ``All'' reports the percentage of predictions having the correct log level (independently from the correct/wrong prediction of location and message); column ``Corr. Pred.'' reports the percentage of completely correct predictions.}
			\begin{tabular}{lccccccccccccccc}
				\toprule
				\multirow{2}{*}{\textbf{Level}} 	&& \multicolumn{2}{c}{\textbf{T5 Multi-Task}} 	&& \multicolumn{2}{c}{\textbf{T5 LogStmt-Task}} 	&& \multicolumn{2}{c}{\textbf{T5 Denoising-Task}}		&& \multicolumn{2}{c}{\textbf{T5 No Pre-training}} \\ 
				\cline{3-4} \cline{6-7} \cline{9-10} \cline{12-13} 
													&& \textbf{All} 		& \textbf{Corr. Pred.}			&&\textbf{All} 		& \textbf{Corr. Pred.} 					&&\textbf{All} 		& \textbf{Corr. Pred.}					&& \textbf{All} 	& \textbf{Corr. Pred.}     	\\
				\midrule
				\textit{Trace} 						&& 54.77\% 				& 7.34\% 				&& 58.15\% 			& 7.48\%     					&& 61.67\% 			& 11.16\% 		  				&& 56.24\%  		& 7.93\% 			\\				\textit{Debug}						&& 61.09\% 				& 13.21\% 				&& 63.26\% 			& 13.34\% 						&& 67.41\% 			& 16.19\% 		   				&& 63.92\% 			& 13.98\%			\\ 
				\textit{Info}						&& 56.53\% 				& 8.00\% 				&& 57.66\% 			& 7.80\%         				&& 60.76\% 			& 10.93\%		  				&& 56.56\% 			& 8.02\%  			\\ 
				\textit{Warn}						&& 54.90\% 				& 7.22\%				&& 56.94\% 			& 7.42\%       					&& 60.20\% 			& 9.59\% 						&& 55.02\% 			& 7.10\%   			\\
				\textit{Error} 						&& 68.73\% 		 		& 20.71\% 				&& 69.69\% 			& 20.56\%  						&& 76.10\% 			& 23.40\% 		   				&& 68.50\% 			& 21.05\%  			\\
				\textit{Fatal}						&& 61.36\% 				& 6.81\% 				&& 65.90\% 			& 13.63\%     					&& 59.09\% 			& 13.63\% 						&& 68.18\% 			& 13.63\% 			\\
				\bottomrule
			\end{tabular}
		\label{tab:logLevel}
\end{table*}

%% file: threats.tex

\eject
\section{Threats to Validity} \label{sec:threats}

We discuss the threats to the validity of our study.

\subsection{Construct validity} 
In our study we use the original code written by developers (in our case, log statements) as oracle, assuming that it represents a good target for our model. This assumption has been made in many previous works applying machine learning on code \cite{Hu:icpc2018,iyer:acl,Allamanis:2016,Hu:emse2020,haque:2020}. However, it is likely that both the training and the evaluation/test datasets contain suboptimal log statements. In terms of training, we expect the DL model to be able to deal with such a noise, not learning unusual logging practices. 

However, when it comes to the evaluation and test set, this assumption can have a strong influence, since we consider a prediction correct only if it is equal to the log statement written by developers. To at least partially address this threat we manually analyzed a sample of wrong predictions, reporting the percentage of them still being valuable while different from the reference.

\subsection{Internal validity} 
We used the default T5 parameters from the original paper \cite{raffel2019exploring} during its pre-training and limited the hyperparameter tuning to the fine-tuning phase and, in particular, to the learning rates (\eg we did not variate the model architecture in terms of number of layers). This was done due to the high cost of training several different models. We acknowledge that experimenting with additional configurations may lead to better results. On top of this, it is worth mentioning that we employed the simplest T5 architecture (\ie the \emph{small} one) proposed by Raffel \etal \cite{raffel2019exploring}. Larger models are likely to push forward the results we achieved.

\subsection{External validity} 
While the datasets used in our study features thousands of instances, we limited our experiments to Java code and, more specifically, to projects relying on the Log4j library. Thus, our results are valid for this specific code population and we do not claim generalizability for other languages and logging frameworks. 

However, excluding the building of the datasets that focused on a specific context, there are no parts of our approach that are customized for Java and/or for the Log4j library. Thus, \tool can be easily adapted and experimented in other contexts.

%% file: related.tex

\section{Related Work} \label{sec:related}

We focus our discussion on (i) empirical studies on logging practices, and (ii) approaches proposed in the literature to support developers in logging activities. Due to lack of space, we do not discuss the many recent applications of deep learning to automate various software engineering tasks, pointing the reader to the systematic literature review by Watson \etal \cite{watson2020systematic}.

\subsection{Empirical Studies on Logging Practices}
Yuan \etal \cite{yuan2012characterizing} conducted one of the first empirical study on logging practices in open-source systems, analyzing C and C++ projects. They show that developers make massive usage of log statements and continuously evolve them with the goal of improving debugging and maintenance activities.

Fu \etal \cite{fu2014developers} studied the logging practices in two industrial projects at Microsoft, investigating in particular which code blocks are typically logged. They also propose a tool to predict the need for a new log statement, reporting a 90\% F-Score.
 
Chen \cite{chen2017characterizing} and Zeng \etal \cite{zeng2019studying} extended the study of Yuan \etal \cite{yuan2012characterizing} to Java and Android systems, respectively. In particular, Chen analyzed 21 Java-based open-source projects while Zeng \etal considered 1,444 open-source Android apps mined from F-Droid. Both studies confirmed the results of Yuan \etal \cite{yuan2012characterizing}, finding a massive presence of log statements in the analyzed systems. 

Zhi \etal \cite{zhi2019exploratory} investigated how logging configurations are used and evolve, distilling 10 findings about practices adopted in logging management, storage, formatting, and configuration quality. Other researchers studied the evolution and stability  of log statements. For example, Kabinna \etal \cite{kabinna2018examining} examined how developers of four open source applications evolve log statements. They found that nearly 20-45\% of log statements change throughout the software lifetime. 

Zhou \etal \cite{zhou2020mobilogleak} explored the impact of logging practices on data leakage in mobile apps. In addition, they propose MobiLogLeak to automatically identify log statements in deployed apps that leak sensitive data. Their study show that 4\% of the analyzed apps leak sensitive data.

Recently, Li \etal~\cite{li2020qualitative} conducted an extensive investigation on logging practice from a developer's perspective. The goal of this research is to push the design of automated tools based on actual developers' needs (rather than on researchers' intuition). 

The authors surveyed 66 developers and analyzed 223 logging-related issue reports shedding light on the trade-off between costs and benefits of logging practices in open source. The results show that developers adopt an \emph{ad hoc} strategy to compensate costs and benefits while inserting logging statements for various activities (\eg debugging). 

The above-described papers lay the empirical foundations for techniques supporting developers in logging activities (including our work). Approaches such as \tool can help in reducing the cost of logging while supporting developers in taking proper decisions when they wish to add log statements.

\subsection{Approaches on Logging Activities}

Researchers proposed techniques and tools to support developers in logging activities.

\textbf{Log message enhancement.} Yuan \etal \cite{yuan2012improving} proposed \textsc{LogEnhancer} as a prototype to automatically recommend relevant variable values for each log statement, refactoring its message to include such values. Their evaluation on eight systems demonstrates that \textsc{LogEnhancer} can dramatically reduce the set of potential root failure causes when inspecting log messages. Liu \etal \cite{liu2019variables} tackled the same problem using, however, a customized deep learning network. Their evaluation showed that the mean average precision of their approach is over 84\%. \smallskip 

\textbf{Log placement.} Other researchers targeted the suggestion of the best code location for log statements \cite{jia2018smartlog,li2018studying,li2020towards}. For example, Zhu \etal \cite{zhu2015learning} presented \textsc{LogAdvisor}, an approach to recommend where to add log statements. The evaluation of \textsc{LogAdvisor} on two Microsoft systems and two open-source projects reported an accuracy of 60\% when applied on pieces of code without log statements.
Yao \etal \cite{yao2018log4perf} tackled the same problem in the specific context of monitoring the CPU usage of web-based systems, showing that their approach helps developers when logging.

Li \etal \cite{li2020shall} proposed a deep learning framework to recommend logging locations at the code block level. They report a 80\% accuracy in suggesting logging locations using within-project training, with slightly worst results (67\%) in a cross-project setting. C\^andido \etal \cite{candido2021exploratory} investigated the effectiveness of log placement techniques in an industrial context. Their findings (\eg 79\% of accuracy) show that models trained on open source code can be effectively used in industry. \smallskip 

\textbf{Log level recommendation.} A third family of techniques focus on recommending the proper log level (\eg error, warning, info) for a given log statement \cite{yuan2012characterizing,oliner2012advances}. Mizouchi \etal \cite{mizouchi2019padla} proposed \textsc{PADLA} as an extension for Apache Log4j framework to automatically change the log level for better record of runtime information in case of anomalies. 
The \textsc{DeepLV} approach proposed by Li \etal \cite{li2021deeplv} uses instead a deep learning model to recommend the level of existing log statements in methods. \textsc{DeepLV} aggregates syntactic and semantic information of the source code and showed its superiority with respect to the state-of-the-art. 
\smallskip 

\tool, as compared to the above discussed techniques, is able to recommend complete log statements and where to inject them, providing a more comprehensive support to software developers.


%% file: conclusion.tex

\section{Conclusion and Future Work} \label{sec:conclusion}
We presented \tool, the first approach able to synthesize complete log statements and inject them in the right code location. \tool is built on top of the Text-To-Text-Transfer-Transformer (T5) model \cite{raffel2019exploring}. We built a dataset composed of $\sim$7M Java methods that have been used for training T5 and testing its performance. We experimented with different pre-training strategies, showing that a simple denoising task (\ie the model is asked to guess masked tokens in Java methods) allows T5 to achieve good performance.

In particular, the best-performing model can generate completely correct log statements and inject them in the proper code location in 15.20\% of cases, with better performance achieved in the simpler tasks of selecting a proper log level (66.24\%) and code location (65.94\%). 

We also showed, through manual inspection of a sample of ``wrong'' predictions, that the results of our quantitative analysis are a lower bound for \tool's performance. Indeed, a non-negligible set ($\sim$28\%) of log statements classified as ``wrong'' due to differences between the generated and the target log message, actually represents valuable recommendations. 

Despite the encouraging results, we acknowledge that \tool is just a first attempt in automatically generating complete log statements and additional improvements are needed before it can be considered as a valid support for developers. This observation guides our future research agenda, that will focus on:

\begin{itemize}
\item \emph{Improving \tool's performance}. This could be achieved in different ways. First, we want to experiment with more complex and multi-modal source code representations that have been shown to boost the performance of DL techniques in code-related tasks \cite{chakraborty2021multimodal}. Second, we plan to consider additional pre-training objectives and to study the role played by the size of the training dataset on \tool's performance. Indeed, it is possible that larger datasets substantially improve performance or that, instead, our dataset was already sufficient for the learning, with its extension only leading to marginal improvements. Finally, we want to enlarge our search space in terms of hyperparameters to optimize the T5 performance.\smallskip

\item \emph{Closing the circle by providing full logging support}. While \tool can generate complete log statements, it cannot decide whether a log statement is needed or not in a given method. This limitation can be addressed in two ways. First, by delegating such a decision to other techniques only in charge of deciding which parts of a system to log \cite{yuan2010sherlog}. Second, by training \tool to also support such a task. This can be done by including in the fine-tuning dataset a mixture of methods featuring and not featuring log statements, asking the model to decide when a log statement is needed. For this first work, we decided to not tackle such a problem due to the many different aspects we needed to explore only for the problem of log statement generation.
\end{itemize} 

\eject

\section{Data Availability}
All the code and data used in our study is publicly available in our replication package \cite{replication}. 
In particular, we provide: (i) the code needed to pre-train, fine-tune, and run the T5 model to generate predictions; (ii) the datasets we built for the model training, evaluation, and testing; (iii) all predictions generated by the different variants of the experimented model; and (iv) additional information needed to replicate our study (\eg the exact values used for the learning rates during hyperparameter tuning).